\documentclass[showpacs,preprintnumbers,twocolumn,amsmath,amssymb, groupedaddress,superscriptaddress]{revtex4-1}
\usepackage{graphicx}
\usepackage{dcolumn}
\usepackage{bm}

\begin{document}

\title{Microscopic analysis of quasielastic scattering and breakup
reactions of neutron-rich nuclei $^{12,14}$Be}

\author{V.~K.~Lukyanov}
\affiliation{Joint Institute for Nuclear Research, Dubna 141980,
Russia}

\author{D.~N.~Kadrev}
\affiliation{Institute for Nuclear Research and Nuclear Energy,
Bulgarian Academy of Sciences, Sofia 1784, Bulgaria}

\author{E.~V.~Zemlyanaya}
\affiliation{Joint Institute for Nuclear Research, Dubna 141980,
Russia}

\author{K.~V.~Lukyanov}
\affiliation{Joint Institute for Nuclear Research, Dubna 141980,
Russia}

\author{A.~N.~Antonov}
\affiliation{Institute for Nuclear Research and Nuclear Energy,
Bulgarian Academy of Sciences, Sofia 1784, Bulgaria}

\author{M.~K.~Gaidarov}
\affiliation{Institute for Nuclear Research and Nuclear Energy,
Bulgarian Academy of Sciences, Sofia 1784, Bulgaria}

\begin{abstract}
A microscopic analysis of the optical potentials (OPs) and cross
sections of quasielastic scattering of $^{12,14}$Be on $^{12}$C at
56 MeV/nucleon and on protons at energy near 700 MeV is carried
out. For lower energy scattering the real part of the OP is
calculated by using of double-folding procedure accounting for the
anti-symmetrization effects, while the imaginary part is obtained
on the base of the high-energy approximation (HEA). The HEA is
also applied to the calculations of both real and imaginary OPs
when solving the relativistic equation for the high-energy
proton-nucleus elastic scattering. The neutron and proton density
distributions computed in different microscopic models for
$^{12}$Be and $^{14}$Be are used. In the present hybrid model of
the optical potential the only free parameters are the depths of
the real and imaginary parts of OP obtained by fitting the
experimental data. The role of the inelastic scattering channel to
the first excited $2^{+}$ and $3^{-}$ states in $^{12}$C when
calculating the quasielastic cross sections, as well as the
modified density of the $^{12}$C target accounting for the surface
effects are studied. In addition, the cluster model, in which
$^{14}$Be consists of a $2n$-halo and the $^{12}$Be core, is
applied to calculate the cross sections of diffraction breakup and
stripping reactions in $^{14}$Be+$^{12}$C scattering and
longitudinal momentum distributions of $^{12}$Be fragments at
energy of 56 MeV/nucleon. A good agreement of the theoretical
results with the available experimental data of both quasielstic
scattering and breakup processes is obtained.
\end{abstract}


\maketitle

\section{Introduction
\label{s:intro}}

Since the pioneering works of Tanihata {\it et. al.}
\cite{Tanihata85a,Tanihata85b} the study of halo nuclei has
attracted much attention. Halo nuclei are commonly considered to
have a compact nuclear core and a few valence nucleons surrounding
the core. Characteristic features displayed by these nuclei
include weak binding energy of the valence nucleons, narrow
momentum distributions of the reaction products due to the
fragmentation and the anomalously large interaction cross section.

The microscopic studies of elastic scattering of $^{6,8}$He,
$^{11}$Li, $^{10,11}$Be, and $^{8}$B on protons and nuclei and
breakup processes performed in our previous works
\cite{Lukyanov2007,Lukyanov2009,Lukyanov2010,Lukyanov2013,Lukyanov2015,Lukyanov2017}
have confirmed the specific internal spatial structure of these
neutron- and proton-halo nuclei and shed light on the relative
contributions of different reaction mechanisms. Studying this
series of light nuclei, the interest of considering very
neutron-rich beryllium isotopes is provoked, for instance, by the
magicity loss for the $N=8$ nucleus $^{12}$Be
\cite{Iwasaki2000a,Iwasaki2000b,Navin2000,Shimoura2003,Pain96} and
the halo structure of $^{14}$Be nucleus, which is located at the
neutron drip line and its two-neutron separation energy is
$S_{2n}=1.26(13)$ MeV \cite{Audi2003}. $^{14}$Be is a Borromean
nucleus like $^{11}$Li and it has a two-neutron halo structure
with a $^{12}$Be core plus two loosely-bound neutrons
\cite{Tanihata88,Zahar93,Suzuki99,Tarutina2004,Gangopadhyay2005}.

Here we note that the task for the structure of the two-neutron
halo nuclei is of important interest in connection with the
general question about the behavior of the dineutron ($2n$)
formations in exotic nuclei. First, we note the work of Migdal
\cite{Migdal72}, in which it was shown that the attractive force
between two neutrons (itself too weak to form a bound $2n$ system)
in the presence of a nucleus (itself unable to bind a single
neutron) may lead to a bound state of the three particles, i.e. it
is a dineutron coupled to a nuclear core (see also studies, e.g.,
in
Refs.~\cite{Hansen87,Hansen93,Otsuka2013,Tostevin98,MacDonald2009,Ilieva2012}).
The possibility that cluster states more complex than dineutrons
may exist has been pointed also in Ref.~\cite{Migdal72}.

The interest to $2n$-formations has increased also in relation to
the experiments that showed a ground state dineutron decay of
$^{16}$Be nucleus \cite{Spirou2012}. It has been observed there a
small angle of emission between two neutrons and a value of the
two-neutron separation energy $S_{2n}=1.35(10)$ MeV has been
measured. Here we note that in the case of the $^{14}$Be nucleus
this energy ($S_{2n}=1.26$ MeV) is close to that in $^{16}$Be.

As noted in Ref.~\cite{Zahar93}, the $^{14}$Be system is even more
interesting than $^{11}$Li since the wave function of the last two
neutrons in $^{14}$Be is expected to contain a larger
$(2s_{1/2})^{2}$ shell-model component. In addition, the
two-neutron separation energy in $^{14}$Be is much larger than
that of $^{11}$Li ($S_{2n}=0.376$ MeV). So, it is of interest to
study this effect of extra binding on the properties of the
neutron halo (see also Ref.~\cite{Abdullah2017}). At the same
time, however, one must bear in mind the relatively small
difference between the halo rms radii of both nuclei (6 fm in
$^{11}$Li \cite{Bertulani93} and 5.5 fm in $^{14}$Be
\cite{Tanihata85a,Tanihata85b,Tanihata88}). All the mentioned
facts give a reason for more detailed studies of these
neutron-rich systems with a $2n$-halo and their interactions with
nuclei.

Many experimental and theoretical studies of the matter density
distributions in nuclei far from stability show an extended
low-density tail at large radial distances in their behavior. As
an example, the calculations in the framework of the relativistic
Hartree-Bogoliubov model have predicted very large neutron skin in
$^{14}$Be and a large prolate deformation of this nucleus
\cite{Lalazissis2004}. Besides, the related with the matter
densities extraordinarily large radii (see, for instance,
Ref.~\cite{Liatard90,Ren90}) are in favor of the halo structure of
the neutron(proton)-rich nuclei.

From the analyses of proton elastic scattering in inverse
kinematics at intermediate energy
 about 700 MeV/nucleon Ilieva {\it et al.} \cite{Ilieva2012} showed an
extended matter distribution for $^{12,14}$Be nuclei. A clear
evidence of a halo structure has been obtained demonstrating
better qualitative description of the $p$--$^{14}$Be cross section
when the $^{14}$Be nucleus is supposed to consist of a $^{12}$Be
core and two halo neutrons rather than a $^{10}$Be core plus four
valence neutrons. Several phenomenological parametrizations
including the symmetrized Fermi function, as well as a sum of
Gaussian ones, were used for the nuclear-matter density
distribution in the analysis performed in Ref.~\cite{Ilieva2012}.

The ground-state proton, neutron, and matter densities, the
corresponding root-mean-square (rms) radii and elastic charge form
factors of $^{12}$Be and $^{14}$Be nuclei have been studied
through shell-model calculations using different model spaces for
the core and the extra two halo neutrons \cite{Radhi2013} and in a
three-body model of (core+$n$+$n$), where the core and halo
density distributions were described by the single-particle wave
functions of the Woods-Saxon (WS) potential \cite{Abdullah2017}. A
renormalized zero-range version of the same three-body model has
been applied to study the rms radii of weakly-bound light nuclei
($^{6}$He, $^{11}$Li, $^{14}$Be, and $^{20}$C), particularly the
mean square distance between the two neutrons forming halo in them
\cite{Yamashita2004}. The good qualitative agreement between the
recently measured data and the theoretical results has indicated
that the model is reasonable for $^{14}$Be validating the large
probability of the halo neutrons to be found outside the
interaction range. Under the assumption of similar decomposition
of the matter density with core and halo contributions, in
Ref.~\cite{Bhagwat2000} simple analytic expressions for nuclear
densities with a correct asymptotic behavior were proposed for
exotic nuclei including the $^{7-14}$Be isotopes.

A "long tail" of neutron density distribution compared with the
proton density distribution in $^{14}$Be nucleus based on the
relativistic mean-field (RMF) theory has been displayed in
Ref.~\cite{Ping2005}. It was shown in Ref.~\cite{Ren95} that the
density-dependent RMF formalism can satisfactorily reproduce the
experimental data of the abnormally large rms radius of $^{14}$Be,
in which the halo neutrons occupy the already mentioned above
$2s_{1/2}$ level instead of the $1d_{5/2}$ level. On the contrary,
the dominance of the $d$-configuration in the $N=8$ shell in
$^{12}$Be was strongly revealed from the breakup reaction on a
proton target at intermediate energy \cite{Chung2017}. Also,
different measurements of reaction cross sections of $^{14}$Be on
protons and carbon target at about 41 and 76 MeV/nucleon
\cite{Moriguchi2014}, on Be, C, and Al targets at several energies
in the range of 45--120 MeV/nucleon \cite{Fukuda2014}, as well as
at relativistic energies \cite{Terashima2014}, allowed one to
deduce the matter density distribution of this two-neutron halo
nucleus supporting the $s$-wave dominance in the ground-state
density of $^{14}$Be. The fact that the ground-state wave function
of $^{14}$Be includes a strong $2s_{1/2}$ admixture has been
confirmed in the experiment of Labiche {\it et al.}
\cite{Labiche2001}, in which they studied the dissociation of
$^{14}$Be at 35 MeV/nucleon on carbon and lead targets in a
kinematically complete measurement. Here we would like to mention
the result for $^{14}$Be nucleus from more sophisticated
microscopic calculations within the three-cluster generator
coordinate method (GCM) \cite{Descouvemont95} involving the ones
for the proton and neutron densities of $^{12,14}$Be.

The widths of the measured momentum distributions following the
fragmentation of $^{12,14}$Be on $^{12}$C at incident energies of
56 and 65 MeV/nucleon have offered a clear qualitative signature
of the spatial distribution of the halo particles
\cite{Zahar93,Kolata94}. The deduced value ($92.2 \pm 2.7$ MeV/c)
of the width parameter of the Lorentzian momentum distribution
that describes the measured $^{12}$Be longitudinal momentum
distribution at 56 MeV/nucleon via the telescope method and the
full width at half maximum (FWHF) equal to $95.6\pm4.2$ MeV/c of
the single Gaussian that fits the distribution at 65 MeV/nucleon
obtained via the spectrograph method were shown to be in agreement
with the "neutron halo" structure of $^{14}$Be. The direct
fragmentation model has been applied in Ref.~\cite{Banerjee95} to
calculate both longitudinal and transverse momentum distributions
of the $^{12}$Be fragments emitted in $^{14}$Be induced breakup
reactions on $^{208}$Pb and $^{12}$C targets at beam energy of 56
MeV/nucleon and the results for the widths are very similar to the
data of Zahar {\it et al.} \cite{Zahar93}.

In the earlier works (e.g., Refs.~\cite{Zahar94,Mermaz94}) the
quasielastic scattering  cross sections of $^{12,14}$Be on
$^{12}$C at 56 MeV/nucleon laboratory incident energy have been
calculated using phenomenological OPs of volume Woods-Saxon shapes
plus surface terms (normalized derivative of WS volume terms) for
both real (ReOP) and imaginary (ImOP) parts. In
Ref.~\cite{Zahar94} for the case of $^{12}$Be+$^{12}$C scattering
such an additional real surface potential was not included. A
substantial difference is seen from the comparison of the values
of the ReOP and ImOP depths in both analyses
\cite{Zahar94,Mermaz94}. For instance, to obtain a good fit of the
experimental $^{12,14}$Be angular distributions the volume real
potentials in \cite{Zahar94} turned out twice deeper than the
corresponding ones shown in Ref.~\cite{Mermaz94}. At the same time
the difference between the values of the volume imaginary
potentials is even larger. This is the reason for the different
total reaction cross sections $\sigma_{R}$ (1238 mb and 1900 mb
for $^{12}$Be and $^{14}$Be projectiles, respectively, in
Ref.~\cite{Zahar94} and 911 mb and 1123 mb in
Ref.~\cite{Mermaz94}). The obtained in Ref.~\cite{Mermaz94} values
of $\sigma_{R}$ fit better the experimentally measured values by
Tanihata {\it et al.} \cite{Tanihata88} (927 mb and 1139 mb,
respectively).

Recently, proximity potentials as an alternative way to produce
the ReOP have been applied in the analysis of scattering cross
sections of Be isotopes \cite{Aygun2018}. However, Woods-Saxon
potential is used for the imaginary part of the OP. The
first-order Dirac OP with direct and exchange parts and
relativistic impulse approximation from Ref.~\cite{Ping2005} have
been applied in Ref.~\cite{Ping2006} to calculate the cross
sections of the elastic scattering of protons at $E_{lab}=100$ and
200 MeV on $^{14}$Be and on stable $^{12}$C and $^{16}$O nuclei.
It has been concluded that the halo neutrons in $^{14}$Be have
effects only in small angular region $4^\circ <\theta < 11^\circ$.
A step ahead in constructing nucleus-nucleus potentials was made
very recently in Ref.~\cite{Durant2018} by using the
double-folding method based on local chiral effective field theory
interactions for the $^{16}$O-$^{16}$O system.

In our present work we aim to perform a fully microscopic analysis
of quasielstic scattering and breakup reactions of neutron-rich
nuclei $^{12,14}$Be. The hybrid model of OP
\cite{Lukyanov2004,Lukyanov2006}, which has been successfully
applied before in our papers
\cite{Lukyanov2007,Lukyanov2009,Lukyanov2010,Lukyanov2013,Lukyanov2015,Lukyanov2017},
is used to analyze the existing data of processes with
$^{12,14}$Be isotopes at incident energies $E<100$ MeV/nucleon
($^{12,14}$Be+$^{12}$C quasielastic scattering)
\cite{Zahar93,Zahar94,Mermaz94} up to relativistic energy of 700
MeV ($p$+$^{12,14}$Be elastic scattering) \cite{Ilieva2012}. In
the folding procedure the ReOP consists of both direct and
exchange potentials with the isoscalar and isovector parts
included. We use the effective nucleon-nucleon potential from
Ref.~\cite{Khoa2000} (see also \cite{Lukyanov2007a}) and
microscopic density distributions for $^{12}$Be obtained within
the variational Monte Carlo (VMC) model \cite{Pieper2015} and the
generator coordinate method \cite{Descouvemont95}. For $^{14}$Be
only the available GCM density \cite{Descouvemont95} is used. The
ImOP is obtained within the HEA model \cite{Glauber,Sitenko},
where the known parametrization of the elementary nucleon-nucleon
($NN$) cross section and scattering amplitude at $\theta=0^\circ$
are used. In contrast to the analyses of quasielastic $^{12,14}$Be
on $^{12}$C performed in Refs.~\cite{Zahar94,Mermaz94} with large
number of optical model fitting parameters, the only free
parameters in our model are the depths of the real and imaginary
parts of the microscopic OP obtained by fitting the experimental
differential cross section data.

We also search for other effects that should be incorporated in
the microscopic study, namely, to account for the inelastic
scattering to the low-lying 2$^{+}$ and 3$^{-}$ collective states
in $^{12}$C in the quasielastic process and the role of the
density distribution of the $^{12}$C target with inclusion of
surface terms. Such an investigation is supposed to figure out the
role of the neutron halo for both Be projectiles. Second, in
addition to the analysis of quasielastic scattering cross
sections, we estimate important characteristics of the reactions
with $^{14}$Be, such as the breakup cross sections for the
diffraction and stripping processes and the momentum distributions
of $^{12}$Be fragments from the breakup reaction
$^{14}$Be+$^{12}$C for which experimental data are available
\cite{Zahar93}. Such a complex study based on the microscopic
method to obtain the OPs with a minimal number of free parameters
and by testing density distributions of $^{12,14}$Be which reflect
their two-neutron halo structure would lead to a better
understanding the structure of these neutron-rich nuclei and to a
reduction of the inconsistency of describing the available data.

The structure of the paper is as follows. The theoretical scheme
to calculate microscopically within the hybrid model the ReOP and
the ImOP, as well as the results for the $^{12,14}$Be+$^{12}$C
quasielastic- and $p$+$^{12,14}$Be elastic-scattering differential
cross sections are presented in Sec.~\ref{s:elastic}.
Section~\ref{s:breakup} contains the basic formulae to estimate
the $^{14}$Be breakup on $^{12}$C in the stripping and diffraction
processes within the cluster model with two-neutrons halo of
$^{14}$Be and the corresponding results for the longitudinal
momentum distributions of $^{12}$Be fragments. The summary and
conclusions of the work are given in Sec.~\ref{s:conclusions}.

\section{Quasielastic scattering of $^{12,14}$Be on $^{12}$C and
protons \label{s:elastic}}
\subsection{Hybrid model for the optical potential\label{s:op}}

The microscopic OP used in our calculations of quasielastic
scattering differential cross sections contains the volume real
part ($V^{F}$) including both the direct and exchange terms and
the HEA microscopically calculated imaginary part ($W^{H}$). It
has the form
\begin{equation}
U(r) = N_R V^{F}(r) + i N_I W^{H}(r).
\label{eq:1}
\end{equation}
The parameters $N_R$ and $N_I$ entering Eq.~(\ref{eq:1})
renormalize the strength of OP and are fitted by comparison with
the experimental cross sections.

The real part $V^{F}$ realized numerically in \cite{Lukyanov2007a}
consists of the direct ($V^{D}$) and exchange ($V^{EX}$)
single(double)-folding integrals that include effective $NN$
potentials and density distribution functions of colliding nuclei.
The $V^{D}$ and $V^{EX}$ parts of the ReOP have isoscalar (IS) and
isovector (IV) contributions. The IS ones of both terms are:
\begin{equation}
V^{D}_{IS}(r) = \int d^3 r_p d^3 r_t  {\rho}_p({\bf r}_p) {\rho}_t
({\bf r}_t) v_{NN}^D(s),
\label{eq:2}
\end{equation}
\begin{eqnarray}
V^{EX}_{IS}(r)&=& \int d^3 r_p d^3 r_t   {\rho}_p({\bf r}_p, {\bf
r}_p+ {\bf s})  {\rho}_t({\bf r}_t, {\bf r}_t-{\bf s})
\nonumber \\
& & \times  v_{NN}^{EX}(s)  \exp\left[ \frac{i{\bf K}(r)\cdot
s}{M}\right],
\label{eq:3}
\end{eqnarray}
where ${\bf s}={\bf r}+{\bf r}_t-{\bf r}_p$ is the vector between
two nucleons, one of which belongs to the projectile and another
one to the target nucleus. In Eq.~(\ref{eq:2}) $\rho_p({\bf r}_p)$
and $\rho_t({\bf r}_t)$ are the densities of the projectile and
the target, respectively, while in Eq.~(\ref{eq:3}) $\rho_p({\bf
r}_p, {\bf r}_p+{\bf s})$ and $\rho_t({\bf r}_t, {\bf r}_t-{\bf
s})$ are the density matrices for the projectile and the target
that are usually taken in an approximate form
\cite{Campy78,Negele72} (see also
Refs.~\cite{Lukyanov2007,Lukyanov2009}). The effective $NN$
interactions $v_{NN}^{D}$ and $v_{NN}^{EX}$ have their IS and IV
components in the form of M3Y interaction obtained within
$g$-matrix calculations using the Paris $NN$ potential
\cite{Khoa2000}. The expressions for the energy and density
dependence of the effective $NN$ interaction are given, e.g., in
Ref.~\cite{Lukyanov2015}. In Eq.~(\ref{eq:3}) ${\bf K}(r)$ is the
local momentum of the nucleus-nucleus relative motion:
\begin{equation}
K(r)=\left \{\frac{2Mm}{\hbar^2}\left[E-V^{F}(r)-V_c(r)\right
]\right \}^{1/2}
\label{eq:4}
\end{equation}
with $M=A_pA_t/(A_p+A_t)$, where $A_{p}$, $A_{t}$, $m$ are the
projectile and target atomic numbers and the nucleon mass. As can
be seen, $K(r)$ depends on the folding potential $V^{F}(r)$ that
has to be calculated itself and, thus, nonlinearity effects occur
as typical ingredients of the model and they have to be taken
carefully into account.

Concerning the ImOP, it corresponds to the full microscopic OP
derived in Refs.~\cite{Lukyanov2004,Lukyanov2006,Shukla2003}
within the HEA \cite{Glauber,Sitenko}:
\begin{eqnarray}
U^H&=&V_H(r)+iW^H(r)= -\frac{\hbar v}{(2\pi)^2}{\bar\sigma}_{N}
(\bar\alpha + i)  \nonumber \\
&&\times \int_0^{\infty} j_0(qr){\rho}_p(q){\rho}_t(q){f}_N(q) q^2dq.
\label{eq:5}
\end{eqnarray}
In Eq.~(\ref{eq:5}) $\rho(q)$ are the corresponding form factors
of the nuclear densities, $\bar\alpha$ is the ratio of the real to
imaginary part of the $NN$ scattering amplitude at forward angles,
$f_N(q)=\exp(-{\bar\beta}q^2/2)$ is the $q$-dependence of the $NN$
scattering amplitude and $\bar\sigma_{N}$ is the total $NN$
scattering cross section that has been parametrized as function of
the energy up to 1 GeV \cite{Charagi92,Xiangzhow98,Grein77}. The
values of $\bar\alpha$, $\bar\sigma_{N}$, and $\bar\beta$ are
averaged over the isospin of the nucleus.

%If one takes into account the in-medium effects then factors from
%\cite{Xiangzhow98,Grein77} should be added to (\ref{eq:5}).

\subsection{Results of calculations of cross sections\label{s:cs}}

We calculate the OP [Eq.~(\ref{eq:1})] and the elastic scattering
cross sections of $^{12,14}$Be on $^{12}$C and protons using the
DWUCK4 code \cite{DWUCK} for solving the Schr\"{o}dinger equation.
All the scattering cross sections will be shown in the figures as
ratios to the Rutherford cross sections $(d\sigma/d\sigma_{R})$.

Concerning the beryllium $^{12,14}$Be isotopes, we apply the
density distributions obtained within the generator coordinate
method \cite{Descouvemont95}. In Ref.~\cite{Descouvemont95} the
$^{14}$Be nucleus is investigated in the three-cluster GCM,
involving several $^{12}$Be+$n$+$n$ configurations. The $^{12}$Be
core nucleus is described in the harmonic-oscillator model with
all possible configurations in the $p$ shell. For the $^{12}$Be
density we use also the one obtained in the framework of the
variational Monte Carlo model \cite{Pieper2015}. In our case,
within the VMC method the proton and neutron densities have been
computed with the AV18+UX Hamiltonian, in which the Argonne v18
two-nucleon and Urbana X three-nucleon potentials are used
\cite{Pieper2015}. Urbana X is intermediate between the Urbana IX
and Illinois-7 models (the latter was used by us in
Ref.~\cite{Lukyanov2015} for the densities of $^{10}$Be nucleus).

Complimentary to both microscopic densities of the neutron-rich
$^{12,14}$Be isotopes, a phenomenological density distribution in
the form of the symmetrized Fermi function (SF) is applied for
them:
\begin{equation}
\rho_{SF}(r)=\rho_{0} \frac{\sinh(R/a)}{\cosh(R/a)+\cosh(r/a)},
\label{eq:6}
\end{equation}
where
\begin{equation}
\rho_{0}=\frac{A}{(4\pi R^{3}/3)}\left[1+\left(\frac{\pi
a}{R}\right )^{2}\right]^{-1}.
\label{eq:7}
\end{equation}
The SF density parameters, the radius $R$ and the diffuseness $a$
in Eq.~(\ref{eq:6}), have been determined in
Ref.~\cite{Ilieva2012} by fitting (within the Glauber approach) to
the experimental cross section data of the $^{12,14}$Be+$p$
elastic scattering at 700 MeV. In our calculations we adopt their
values, namely $R=1.37$ fm, $a=0.67$ fm for $^{12}$Be and $R=0.99$
fm, $a=0.84$ fm for $^{14}$Be. Here we would like to note the
bigger diffuseness parameter $a$ in the case of $^{14}$Be nucleus,
which supports the existence of a halo structure in it. The same
SF form with radius and diffuseness parameters 2.275 fm and 0.393
fm was taken for the density of $^{12}$C target nucleus when
calculating the OPs for $^{12,14}$Be+$^{12}$C quasielastic
scattering.

Additionally, we apply a modified SF density of $^{12}$C
\begin{equation}
\rho(r)=\rho_{SF}(r)+\rho_{SF}^{(1)}(r),
\label{eq:8}
\end{equation}
where the surface effects are revealed through the term
$\rho_{SF}^{(1)}(r)$ being the first derivative of $\rho_{SF}(r)$.
The parameters of this density were obtained in
Ref.~\cite{Burov98} by fitting to electron-nucleus scattering
data. In general, the form (\ref{eq:8}) of the density
distribution has a specific bump near the nuclear surface, where
the elastic process is expected mainly to take place.

\begin{figure}
\centering
\includegraphics[width=\linewidth]{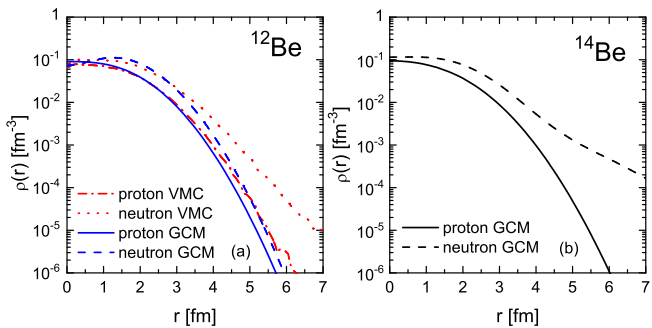}
\caption[]{(a) Point-proton (normalized to $Z=4$) and
point-neutron (normalized to $N=8$) densities of $^{12}$Be
obtained in the VMC method and in the GCM; (b) Point-proton
(normalized to $Z=4$) and point-neutron (normalized to $N=10$)
densities of $^{14}$Be obtained in the GCM.
\label{fig1}}
\end{figure}

As can be seen from Fig.~\ref{fig1}, the proton densities are very
similar in $^{12}$Be and $^{14}$Be nuclei. On the contrary,
neutron densities are quite different: whereas neutron density in
$^{12}$Be is nearly proportional to the proton density, the
neutron contribution in $^{14}$Be has a very long tail. This
long-range neutron density is typical for neutron-rich halo nuclei
and yields fairly large rms radii (value of 2.95 fm obtained in
GCM was reported in Ref.~\cite{Descouvemont95}). One can observe
also from Fig.~\ref{fig1} a different behavior of the
point-neutron densities of $^{12}$Be calculated with GCM and VMC
method. In the GCM the $^{12}$Be internal wave functions are
defined in the $p$-shell harmonic-oscillator model that leads to a
more steep decrease of the corresponding density
\cite{Descouvemont95}. On the contrary, the VMC neutron density
exhibits a broader shape, presumably due to the $^{10}$Be core
plus $2n$ cluster structure effectively accounted for in the
variational calculations \cite{Pieper2015}. Similar behavior of
the proton and neutron densities of $^{10}$Be and $^{11}$Be nuclei
obtained in the GCM and in the quantum Monte Carlo method can be
seen from Fig.~1 of Ref.~\cite{Lukyanov2015}. As a result from
these differences in the surface region, the neutron rms radius
$r_{n}$ of $^{12}$Be obtained within the VMC method has a value of
2.60 fm that is larger than the corresponding value of $r_{n}$
deduced from the GCM (2.33 fm).

\begin{figure}
\centering
\includegraphics[width=\linewidth]{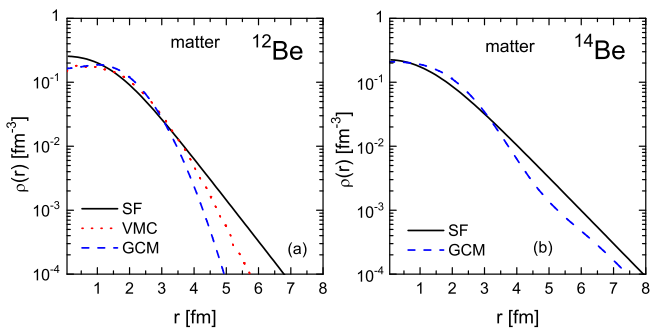}
\caption[]{(a) Matter density distribution (normalized to $A=12$)
of $^{12}$Be obtained with SF function, in the VMC method and in
the GCM; (b) Matter density distribution (normalized to $A=14$) of
$^{14}$Be obtained with SF function and in the GCM.
\label{fig2}}
\end{figure}

In Fig.~\ref{fig2} we present the matter density distributions in
$^{12}$Be and $^{14}$Be nuclei. As can be seen, the SF matter
density of $^{12}$Be exceeds the VMC and GCM densities in the
central region ($r<1.5$ fm), while in the region $2<r<3$ fm its
values are smaller than the ones of the two microscopic densities,
which signals for a mixed $p-sd$ state for the valence neutrons in
$^{12}$Be \cite{Ilieva2012}. Also, the SF matter density of
$^{12}$Be indicates an extended tail in comparison with VMC and
GCM densities. It was mentioned in Ref.~\cite{Ilieva2012} that
there is a tendency for a slightly larger rms matter radius, as
compared to those obtained in previous measurements. In addition,
the relatively big diffuseness parameter $a=0.67$ fm obtained for
the SF model leads to the enhanced matter distribution in the
$^{12}$Be nucleus.

The values of the rms radii of the point-proton, point-neutron,
and matter distributions of $^{12,14}$Be used in our calculations
are listed in Table~\ref{tab1} together with the experimental data
deduced from the Glauber analysis of the interaction and reaction
cross sections \cite {Tanihata88}. In addition, the values of the
matter rms radii of $^{12}$Be and $^{14}$Be nuclei of SF
distributions shown in Fig.~\ref{fig2} are 2.71 fm and 3.22 fm,
correspondingly \cite{Ilieva2012}.

\begin{table}
\caption{Proton, neutron, and matter rms radii (in fm) of
$^{12}$Be and $^{14}$Be nuclei obtained within the VMC method
\cite{Pieper2015} and GCM \cite{Descouvemont95}. Experimental data
are taken from Ref.~\cite{Tanihata88}. \label{tab1}}
\begin{center}
\begin{tabular}{ccccc}
\hline \hline
Nucleus &  Model & $r_{p}$  & $r_{n}$ & $r_{m}$ \\
\hline
$^{12}$Be &  VMC                    & 2.29 & 2.60 & 2.50 \\
          &  GCM                    & 2.20 & 2.33 & 2.29 \\
          &  Exp.~\cite{Tanihata88} & 2.49 & 2.65 & 2.59 \\
\hline
$^{14}$Be &  GCM                    & 2.28 & 2.95 & 2.78 \\
          &  Exp.~\cite{Tanihata88} & 3.00 & 3.22 & 3.16 \\
\hline \hline
\end{tabular}
\end{center}
\end{table}
\begin{figure}
\centering
\includegraphics[width=0.9\linewidth]{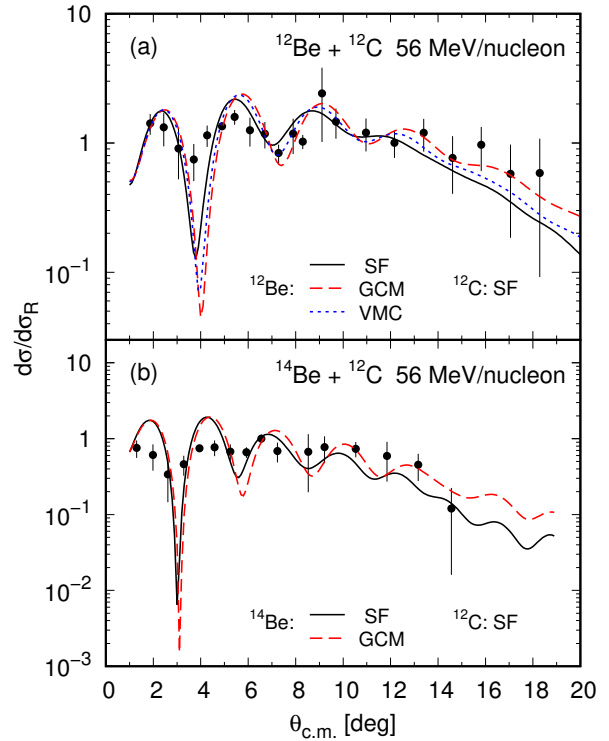}
\caption[]{$^{12}$Be+$^{12}$C (a) and $^{14}$Be+$^{12}$C (b) at
$E=56$ MeV/nucleon elastic scattering cross sections. Black solid
line: calculations with the SF densities of $^{12,14}$Be; red
dashed line: calculations with the GCM densities of $^{12,14}$Be;
blue dotted line: calculations with the VMC density of $^{12}$Be.
Experimental data are taken from Ref.~\cite{Zahar94}.
\label{fig3}}
\end{figure}

\subsubsection{Quasielastic scattering cross sections of
$^{12,14}$Be+$^{12}$C}

Similarly to our previous works (for instance,
Ref.~\cite{Lukyanov2017}), we consider the set of the $N_{i}$
coefficients ($N_R$ and $N_I$, see Eq.~(\ref{eq:1}) for the OP) as
parameters to be found out from the fit to the experimental data
for the cross sections using the $\chi^2$-procedure. The fitted
$N$s related to the depths of the ReOP and ImOP can be considered
as a measure of deviations of our microscopic OPs from the case
when the values of $N$s are equal to unity.

It is worth to mention that the experimental data of $^{12,14}$Be
scattering on $^{12}$C \cite{Zahar94,Mermaz94} are considered to
include contributions of the scattering to the first excited
$2^{+}$ (4.439 MeV) and $3^{-}$ (9.641 MeV) states of $^{12}$C.
Therefore, to calculate the angular distributions and to compare
them with the experimental data we write the following sum:
\begin{equation}
\left (\frac{d\sigma}{d\sigma_{R}}\right )_{quasi}=\left
(\frac{d\sigma}{d\sigma_{R}}\right )_{el}+C\left
(\frac{d\sigma}{d\sigma_{R}}\right )_{inel}, \label{eq:9}
\end{equation}
where the first term corresponds to the pure elastic scattering,
while the second term gives the contribution of the inelastic
scattering to the $2^{+}$ and $3^{-}$ states of $^{12}$C. The
account for the latter states is important since the coupling
between $2^{+}$ state and the ground state of $^{12}$C is strong.
In Eq.~(\ref{eq:9}) the coefficient $C$, which is an additional
fitting parameter, is related with the potential radius $R_{pot}$
and the deformation parameter $\beta$ as $C=(\beta R_{pot})^{2}$.
We adopt $R_{pot}=4.25$ fm as in Ref.~\cite{Zahar94}. Then, the
value of the parameter $\beta$ can be determined. Concerning the
contribution of the inelastic channel, the inelastic OP is
calculated within our approach via the microscopic optical
potential (\ref{eq:1}): $U_{inel}(r)=-R_{pot}[dU(r)/dr]$.

First, before estimating the role of the inelastic channels in the
scattering process it is useful to perform calculations of the
elastic scattering only. The calculated within the hybrid model
elastic scattering cross sections of $^{12}$Be+$^{12}$C and
$^{14}$Be+$^{12}$C at energy $E=56$ MeV/nucleon in the laboratory
frame are given in Fig.~\ref{fig3} and compared with the
experimental data \cite{Zahar94}. It can be seen in the case of
the $^{12}$Be+$^{12}$C scattering that, with the exception of the
deep first minimum, all three SF, GCM, and VMC densities of
$^{12}$Be give a reasonable agreement with the data. In the case
of $^{14}$Be+$^{12}$C, however, an agreement can be seen only at
$\theta_{c.m.} > 5^\circ$.

\begin{figure}
\centering
\includegraphics[width=0.9\linewidth]{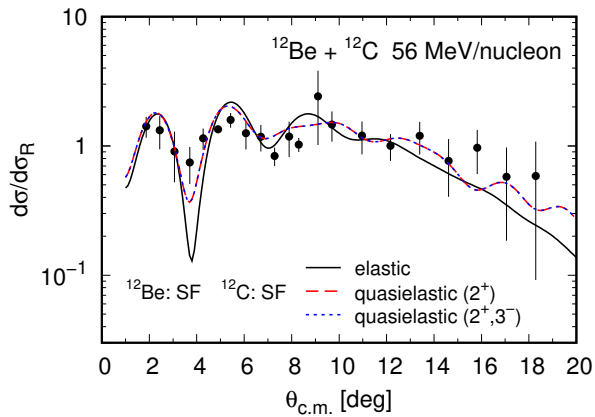}
\caption[]{$^{12}$Be+$^{12}$C quasielastic scattering cross
sections at $E=56$ MeV/nucleon calculated using the SF density of
$^{12}$Be. Black solid line: pure elastic scattering; red dashed
line: elastic plus inelastic scattering to the $2^{+}$ state of
$^{12}$C; blue dotted line: elastic plus inelastic scattering to
the first $2^{+}$ and $3^{-}$ states of $^{12}$C. Experimental
data are taken from Ref.~\cite{Zahar94}.
\label{fig4}}
\end{figure}

\begin{figure}
\centering
\includegraphics[width=0.9\linewidth]{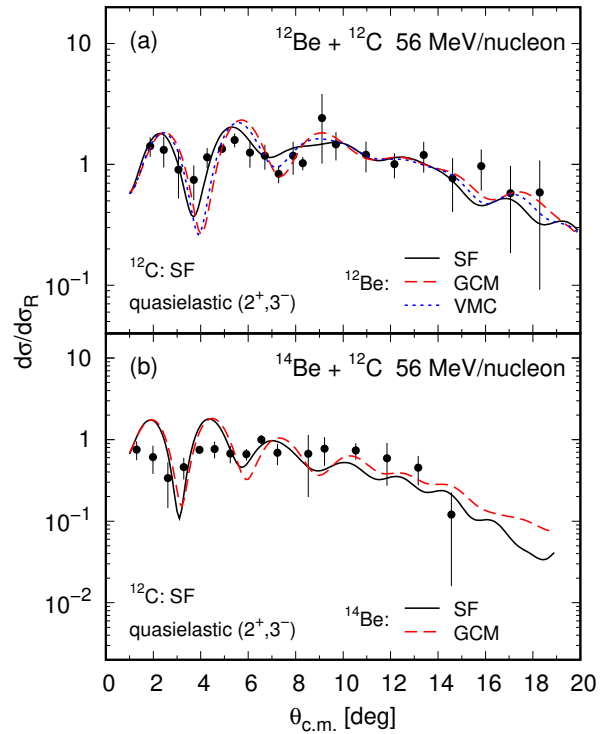}
\caption[]{The same as in Fig.~\ref{fig3} but for the quasielastic
scattering cross sections accounting for inelastic scattering to
the first $2^{+}$ and $3^{-}$ states in $^{12}$C.
\label{fig5}}
\end{figure}

Here we note that the experimental data given in Refs.
\cite{Zahar94,Mermaz94} are presented by the authors as
quasielastic cross sections, in which there are contributions of
elastic and also of inelastic scattering with an excitation of
low-lying 2$^+$ and 3$^-$ states of $^{12}$C nucleus. In
Fig.~\ref{fig4} we give the cross section for the quasielastic
$^{12}$Be+$^{12}$C process using only the SF density and including
the contribution of the inelastic scattering to first 2$^+$ and
3$^-$ states. It can be seen that the account for the inelastic
scattering reduces the depth of the first minimum and provides the
left-shift correction of its place. We note that the role of the
scattering to the 3$^-$ state turns out to be negligible. The
similar qualities of the results can be seen in Fig.~\ref{fig5}
where the quasielastic cross sections calculated with the SF, GCM,
and VMC densities of $^{12}$Be for the $^{12}$Be+$^{12}$C case and
with SF and GCM densities of $^{14}$Be for the $^{14}$Be+$^{12}$C
case are considered.

\begin{figure}
\centering
\includegraphics[width=0.9\linewidth]{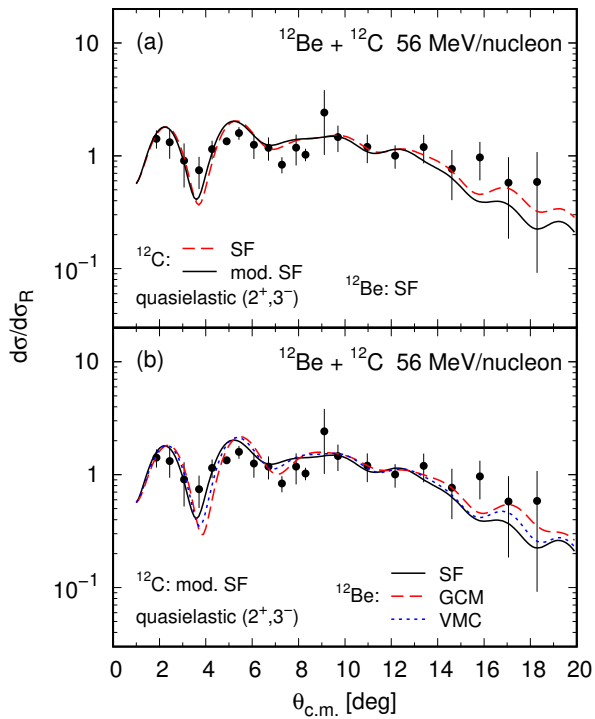}
\caption[]{(a) $^{12}$Be+$^{12}$C quasielastic scattering cross
sections at $E=56$ MeV/nucleon calculated using the SF density of
$^{12}$Be and SF [Eq.~(\ref{eq:6})] (black solid line) and
modified SF [Eq.~(\ref{eq:8})] (red dashed line) densities of
$^{12}$C; The panel (b) illustrates the calculations with the
modified SF density of $^{12}$C [Eq.~(\ref{eq:8})] and using the
SF (black solid line), GCM (red dashed line), and VMC (blue dotted
line) densities of $^{12}$Be. \label{fig6}}
\end{figure}

In the upper part of Fig.~\ref{fig6} we present the quasielastic
cross section for the $^{12}$Be+$^{12}$C case using the SF density
of $^{12}$Be with excitation of 2$^+$ and 3$^-$ states of $^{12}$C
and also including the surface part ($\rho^{(1)}_{SF})$ of the
$^{12}$C density. The latter leads to a further decrease of the
depth of the first minimum. As can be seen in the lower part of
Fig.~\ref{fig6} the situation is similar also in the cases of GCM
and VMC densities of $^{12}$Be. In our opinion, the use of the SF
density gives a better agreement with the data.

\begin{figure}
\centering
\includegraphics[width=0.9\linewidth]{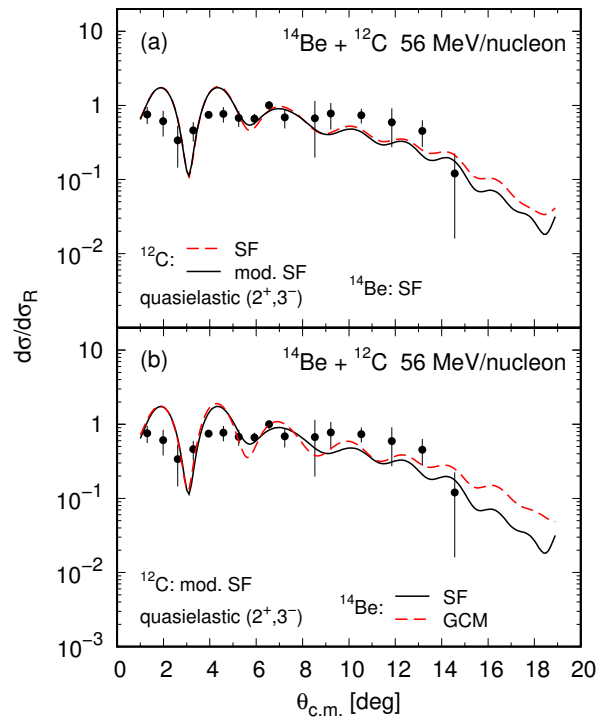}
\caption[]{The same as in Fig.~\ref{fig6} but for the
$^{14}$Be+$^{12}$C quasielastic scattering cross sections at
$E=56$ MeV/nucleon. In panel (b) results with SF and GCM densities
of $^{14}$Be are shown. \label{fig7}}
\end{figure}

The results of the calculations in the case of $^{14}$Be+$^{12}$C
at 56 MeV/nucleon given in Fig.~\ref{fig7} show that the account
for the surface part ($\rho^{(1)}_{SF}$) of the $^{12}$C density
does not improve the agreement of the quasielastic scattering at
angles $\theta_{c.m.} < 5^\circ$ for both SF and GCM densities. As
can be seen from Figs.~\ref{fig4}-\ref{fig7} a better agreement
with the data (in the case of $^{12}$Be+$^{12}$C) up to $8^\circ$
is obtained by accounting only for the elastic channel, while for
larger angles up to $17^\circ$ the contributions of the elastic
and inelastic scattering (with an excitation mainly of the 2$^{+}$
state) are similar in their magnitude and their sum gives an
agreement with the data.

The obtained values of the parameters $N_R$, $N_I$, the
deformation parameter $\beta_{2^+}$, and the total reaction cross
section $\sigma_R$ for the $^{12,14}$Be+$^{12}$C quasielastic
scattering at 56 MeV/nucleon incident energy are presented in
Table~\ref{tab2} for the different densities, for the pure elastic
channel, also when the inelastic channels are included, and when
the surface part of the target $^{12}$C density is accounted for.
It can be seen from Table~\ref{tab2} that our "best fit" results
to the experimental angular distributions using microscopic OPs
lead to values of the predicted total reaction cross sections
$\sigma_{R}$ occupying an intermediate region between the
respective values (discussed in the Introduction) from the
analyses of the data in Refs.~\cite{Zahar94} and \cite{Mermaz94}.

\subsubsection{Elastic scattering cross sections of
$^{12,14}$Be+$p$}

\begin{figure}
\centering
\includegraphics[width=0.9\linewidth]{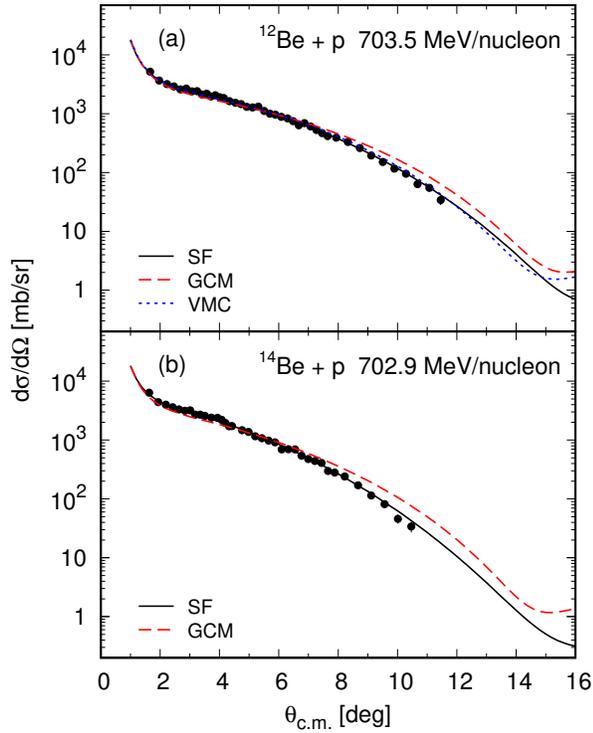}
\caption[]{Differential cross sections for $^{12}$Be+$p$ at
$E=703.5$ MeV/nucleon (a) and $^{14}$Be+$p$ at $E=702.9$
MeV/nucleon (b) elastic scattering. Calculations are performed
with the modified SF density of $^{12}$C [Eq.~(\ref{eq:8})] and
different densities of $^{12,14}$Be. Black solid line:
calculations with the SF densities of $^{12,14}$Be; red dashed
line: calculations with the GCM densities of $^{12,14}$Be; blue
dotted line: calculations with the VMC density of $^{12}$Be.
Experimental data are taken from Ref.~\cite{Ilieva2012}.
\label{fig8}}
\end{figure}

In Fig.~\ref{fig8} we present, in comparison with the experimental
data from \cite{Ilieva2012}, our results of calculations for the
cross sections of $^{12}$Be+$p$ scattering at $E= 703.5$
MeV/nucleon (upper panel) and of $^{14}$Be+$p$ at $E= 702.9$
MeV/nucleon (lower panel) using SF, GCM, and VMC densities in the
former case and SF and GCM densities in the latter case.

As shown in \cite{Lukyanov15,Lukyanov19}, the effects of
relativization are very important at these energies. Here in
calculating differential cross sections the respective optical
potentials (\ref{eq:5}) are used dependent on the relativistic
velocity $v=k/\sqrt{k^2+m^2}$ ($c=1$) for high energies. For our
purposes the DWUCK4 code \cite{DWUCK} was adapted for relativistic
energies to solve the relativistic wave equation at kinetic
energies $T \gg |U^H|$ (below $\hbar$=$c$=1):
%(6)
\begin{equation}
\left(\Delta + k^2 \right) \psi({\mathbf{r}}) = 2 \bar\mu U (r)
\psi({\mathbf{r}}), \quad U = U^H + U_C.
\label{eq:10}
\end{equation}
In Eq.~(\ref{eq:10}) $k$ is the relativistic momentum of a nucleon
in center-of-mass (c.m.) system,
%(7)
\begin{equation}
k= \frac{M k^{lab}}{\sqrt{(M+m)^2+2MT^{lab}}},\quad
k^{lab}= \sqrt{ T ^{lab} \left( T^{lab} + 2m\right)},
\end{equation}
$\bar\mu=E M/(E+M)$ is the relativistic reduced mass with $E =
\sqrt{k^2+m^2}$ being the total energy in c.m. system, $T^{lab}$
is the kinetic energy, $m$ and $M$ are the nucleon and nucleus
masses.

In Ref.~\cite{Lukyanov19} the  $^{12,14}$Be+$p$ cross sections at
700 MeV/nucleon were calculated by using the $NN$ amplitude
parameters ${\bar\sigma}_N$ and $\bar\alpha$ from
Ref.~\cite{Ilieva2012}, and they reasonably reproduce the
experimental data. In this work, to improve the agreement with the
data \cite{Ilieva2012}, the respective parameters ${\bar\sigma}_N$
and $\bar\alpha$ of the optical potential (\ref{eq:5}) were fitted
to the data and the obtained results are presented in
Table~\ref{tab3}. The value of $\bar\beta = 0.17$ fm$^2$ from
Ref.~\cite{Ilieva2012} was used.

It can be seen from Fig.~\ref{fig8} that for $^{12}$Be+$p$
scattering the tested VMC density provides a reasonable agreement
with all the data, and the GSM density is consistent with the data
only at $\theta < 8^\circ$, while in the case of SF density one
gets good fit of the data at all angles of scattering. In the case
of $^{14}$Be one obtains the remarkable accordance with the data
for the SF density, while in the case of GCM density a
considerable excess of the data at $\theta > 7^\circ$ is seen.

\begin{table}
\caption{The renormalization parameters $N_{R}$, $N_{I}$, the
deformation parameter $\beta_{2^{+}}$, and the total reaction
cross sections $\sigma_{R}$ (in mb) for results of the
$^{12,14}$Be+$^{12}$C quasielastic scattering processes at 56
MeV/nucleon incident energy considered and shown in
Figs.~\ref{fig3}--\ref{fig7} using different model densities of
$^{12,14}$Be and $^{12}$C (for details, see the text).
\label{tab2}}
\begin{center}
\begin{tabular}{ccccccc}
\hline \hline
Nucleus &  Model & Model  & $N_R$ & $N_I$ & $\beta_{2^{+}}$  &$\sigma_R$ \\
        &$^{12,14}$Be  &$^{12}$C&      &       &        &      \\
\hline
$^{12}$Be elastic&     SF   &  SF     & 0.767 & 0.593 &       & 1124.80 \\
  &     GCM  &         & 0.804 & 0.855 &       & 1018.47 \\
         &     VMC  &         & 0.721 & 0.660 &       & 1055.15 \\
%channel  &     VMC  &         & 0.721 & 0.660 &       & 1055.15 \\
\hline
quasielastic&     SF   &         & 0.702 & 1.294 & 0.635 & 1353.04 \\
            &     GCM  &         & 0.496 & 1.431 & 0.437 & 1111.34 \\
            &     VMC  &         & 0.583 & 1.156 & 0.487 & 1180.38 \\
\hline
         &     SF   & mod. SF & 0.647 & 1.094 & 0.665 & 1422.01 \\
         &     GCM  &         & 0.592 & 1.133 & 0.526 & 1228.79 \\
         &     VMC  &         & 0.596 & 1.106 & 0.593 & 1309.78 \\
\hline
$^{14}$Be elastic&     SF   & SF      & 0.913 & 1.310 &       & 1666.02 \\
  &     GCM  &         & 1.080 & 2.000 &       & 1597.48 \\
%         &          &         &       &       &       &         \\
\hline
quasielastic&     SF   &         & 0.701 & 1.252 & 0.365 & 1636.39 \\
            &     GCM  &         & 0.638 & 2.000 & 0.375 & 1583.53 \\
\hline
         &     SF   & mod. SF & 0.599 & 0.952 & 0.362 & 1629.13 \\
         &     GCM  &         & 0.708 & 1.920 & 0.369 & 1701.64 \\
\hline \hline
\end{tabular}
\end{center}
\end{table}
\begin{table}
\caption{Parameters $\bar\sigma_N$ (in fm$^2$), $\bar\alpha$, and
the total reaction cross sections $\sigma_{R}$ (in mb) for results
of the $^{12,14}$Be+$p$ elastic scattering processes at incident
energies $E=702.9$ and $E=703.5$ MeV/nucleon considered and shown
in Fig.~\ref{fig8} using different model densities of
$^{12,14}$Be.} \label{tab3}
\begin{center}
\begin{tabular}{ccccccc}
\hline \hline
Nucleus & $E/A$    &  Model & $\bar\sigma_N$ & $\bar\alpha$ &$\sigma_R$& \\
\hline
$^{12}$Be& 703.5   &  SF    & 4.4 & -0.237 &  278.49 \\
        &          &  GCM   & 3.5 & -0.483 &  219.02 \\
        &          &  VMC   & 3.8 & -0.416 &  246.18 \\
\hline
$^{14}$Be& 702.9   &  SF    & 4.136 &-0.2086 &  333.19 \\
        &          &  GCM   & 3.46  & -0.35  &  270.24 \\
\hline \hline
\end{tabular}
\end{center}
\end{table}

\section{Breakup reactions of $^{14}$Be \label{s:breakup}}

Along with the small separation energy of one- or two-neutrons
(protons) and a large rms radius of the corresponding density
distribution, the narrow momentum distributions of the fragments
in the breakup of a given nucleus is a proof of a largely extended
distribution. In our previous works we calculated the breakup
cross sections and momentum distributions of the cluster fragments
in the scattering of $^{11}$Li on protons at 62 MeV/nucleon
\cite{Lukyanov2013}, of $^{11}$Be on $^{9}$Be, $^{93}$Nb,
$^{181}$Ta, and $^{238}$U \cite{Lukyanov2015}, as well as of
$^{8}$B on $^{9}$Be, $^{12}$C, and $^{197}$Au targets
\cite{Lukyanov2017}. A cluster model, in which the nucleus
consists of a halo and a core, has been used in the calculations.
In the present section we calculate the breakup cross sections and
momentum distributions of $^{12}$Be fragments from the breakup of
the halo-nucleus $^{14}$Be on $^{12}$C at energy 56.8 MeV/nucleon
\cite{Zahar93}. This part of the work is related to the already
mentioned in the Introduction general question about the behavior
of dineutron formations in exotic nuclei predicted theoretically
in Ref.~\cite{Migdal72} and considered also, e.g., in
Refs.~\cite{Hansen87,Hansen93,Otsuka2013,Tostevin98,MacDonald2009,Ilieva2012}),
as well as to the results of the experiments on dineutron decay of
$^{16}$Be and its properties observed in Ref.~\cite{Spirou2012}.

\subsection{The $^{12}$Be+$2n$ model of $^{14}$Be}

We consider the characteristics of breakup processes of $^{14}$Be
nucleus, namely diffraction and stripping reaction cross sections
and momentum distributions of the fragments. A simple cluster
model in which $^{14}$Be consists of $^{12}$Be core and a valence
$2n$-halo is used. In this case the hybrid model is applied to
calculate the OPs of the interactions of $^{12}$Be and $2n$ with
the target. The sum of OPs is folded with the density distribution
which corresponds to the wave function of the relative motion of
the clusters in $^{14}$Be. This wave function is obtained as a
solution of the Schr\"{o}dinger wave equation with the WS
potential for a particle with a reduced mass of the two clusters.
The values of the parameters of the WS potentials are obtained by
a fitting procedure to reach the empirical two-neutron separation
energy $S_{2n}$ of the dineutron halo and the rms radius $R_{rms}$
corresponding to the cluster wave function.

The eikonal formalism for the $S$-matrix as a function of the
impact parameter $b$ is used to calculate the breakup cross
sections and momentum distributions of fragments:
\begin{equation}
S(b)=\exp\left [{-\frac{i}{\hbar v}\int_{-\infty}^{\infty}
U(\sqrt{b^{2}+z^{2}})dz}\right ], \label{eq:S}
\end{equation}
where
\begin{equation}
U=V+i W
\label{eq:OP}
\end{equation}
is the OP.
The probability that after the collision with the target
($z\rightarrow \infty$) the core ($c$) or the valence halo ($v=2n$) with
impact parameter $b$ remains in the elastic channel ($i=c,v$) is
given by:
\begin{equation}
|S_{i}(b)|^{2}=\exp{\left[-\frac{2}{\hbar
v}\int_{-\infty}^{\infty} dz\,\left |W_i
(\sqrt{b^{2}+z^{2}})\right |\right ]}.
\label{eq:prob}
\end{equation}
The probability of a cluster to be removed from the elastic
channel is $(1-|S_{i}|^{2})$. The probability both clusters ($c$
and $v$) to leave the elastic channel is
$(1-|S_{c}|^{2})(1-|S_{v}|^{2})$. We note that this procedure can
lead to several groups of parameters of the OPs which fulfill the
conditions. They can be similar, but at the same time to lead to
different values of the rms radius of the $2n$-cluster (the
distance between $^{12}$Be and $2n$ in the case of $^{14}$Be). In
our calculations we use different values of the rms radii
corresponding to the cluster wave functions, like what has been
done in our work on $^{8}$B breakup processes (see
Ref.~\cite{Lukyanov2017}, Table~4), where we used three values of
the relative distances in the system of $^{7}$Be and $p$ clusters.

For the cross section of the breakup of the incident nucleus ($ a $)
into two clusters ($ a+A \rightarrow c+v+A $)
we use, following Ref.~\cite{Hencken96}, the form:
\begin{widetext}
\begin{eqnarray}
\frac{d\sigma}{dk_{\parallel}dk_{\perp}}&=&
\frac{1}{2l+1}\frac{4k_{\perp}}{k^2} \int d^2b \sum_{M,m}
\left |\int dr \int d(\cos\theta) \sum_{L} (-i)^L u_{k,L}(r) g_l(r)
\tilde Y_{L,M}(\theta_k) \tilde Y^*_{L,M}(\theta) \tilde Y_{l,m}(\theta)
\right. \nonumber\\
& &\times \left. \int d\varphi \exp\left(i (m-M)\varphi\right) S_c(\mathbf{b}_c)
S_v(\mathbf{b}_v) \right |^{2},
\label{eq:cs}
\end{eqnarray}
\end{widetext}
where ${\bf k}$ is the relative momentum of both clusters in their
c.m. frame and $ k_{\parallel} $ and $ k_{\perp} $ are its
parallel and transfers components. The relative motion wave
function of the fragments of $a=c+v$ in the continuous final state
was taken as
\begin{equation}
\phi_{{\bf k}}({\bf r})=4\pi\sum_{L,M} i^L \frac{u_{k,L}(r)}{kr}Y_{LM}(\hat r)Y^*_{LM}(\hat k),
\end{equation}
where in the further estimations we neglect the distortion effect and thus use $u_{k,L}(r)=krj_L(kr)$.
Also, $g_l(r)$ is the radial part of the initial bound state wave function of the clusters $c$ and $v$ and
$Y_{LM}(\hat k)=\tilde Y_{L,M}(\theta_k) \exp (iM\varphi_k)$.

In the case of the $s$-state for the mutual motion of the clusters in the incident
nucleus $a=c+v$ the cross sections of the stripping reaction when the valance cluster
$v=2n$ leaves the elastic channel is:
\begin{eqnarray}
\left(\frac{d\sigma}{dk_{\parallel}}\right)_{\text{str}}&=&\frac{1}{2\pi^{2}}\int%_{0}^{\infty}
d^2b_v
\left [ 1-|S_{v}(b_{v})|^{2}\right ]
\int d^2\rho |S_{c}(b_{c})|^{2}
\nonumber \\
&& \times \left [ \int
dz \cos (k_{\parallel}z)\phi_{0}\left
(\sqrt{\rho^{2}+z^{2}}\right )\right]^{2},
\label{eq:strip}
\end{eqnarray}
with ${\bf r}={\mathbf \rho} + {\bf z}$ and ${\bf \rho}={\bf b}_{v}-{\bf b}_{c}$.

\subsection{Results of calculations of breakup reactions}

\begin{figure}
\centering
\includegraphics[width=0.9\linewidth]{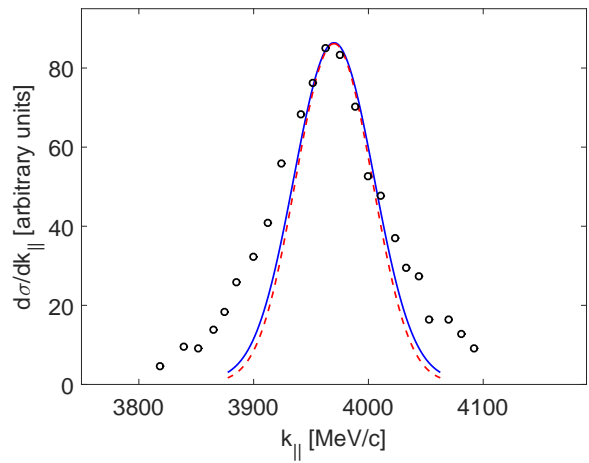}
\caption{Cross sections of  stripping reaction in
$^{14}$Be+$^{12}$C scattering at 56.8 MeV/nucleon. Blue solid
line: result with rms radius $R_{rms}=3.10$ fm, red dashed line:
result with rms radius $R_{rms}=3.50$ fm. Experimental data are
taken from Fig~3(a) of Ref.~\cite{Zahar93}. \label{fig9}}
\end{figure}

\begin{figure}
\centering
\includegraphics[width=0.9\linewidth]{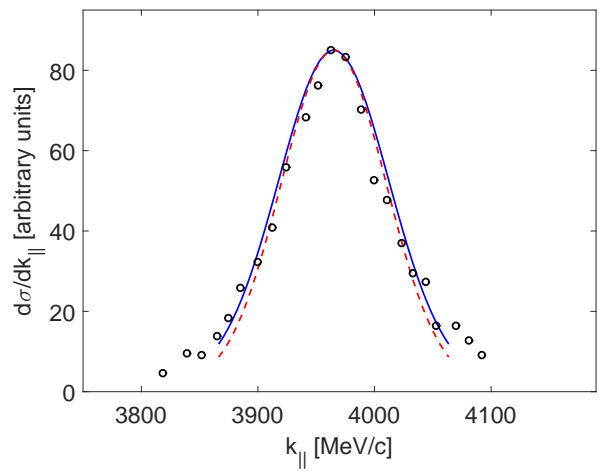}
\caption{Cross sections of diffraction breakup reaction in
$^{14}$Be+$^{12}$C scattering at 56.8 MeV/nucleon. Blue solid
line: result with rms radius $R_{rms}=3.10$ fm, red dashed line:
result with rms radius $R_{rms}=3.50$ fm. Experimental data are
taken from Fig~3(a) of Ref.~\cite{Zahar93}. \label{fig10}}
\end{figure}

The results of our calculations of the $^{12}$Be longitudinal
momentum distribution from $^{14}$Be fragmentation on $^{12}$C at
incident energy of 56.8 MeV/nucleon for stripping and diffraction
processes are given in Figs.~\ref{fig9} and \ref{fig10},
correspondingly. In both figures they are compared with the
experimental data taken from Fig.~3(a) of Ref.~\cite{Zahar93}
(obtained there via the telescope method). In order to check the
sensitivity of the results towards the value of the rms radius
corresponding to the wave function of the relative motion of the
clusters in $^{14}$Be, in each of the figures we present two
theoretical curves. They illustrate the results for rms radii
$R_{rms}=3.10$ fm and $R_{rms}=3.50$ fm which are related to
possible estimated limits of the values of rms radius of
$^{14}$Be, namely, when using its total SF density ($r_{m}=3.22$
fm) and also by estimations on the base of the values of the rms
radius of the core $^{12}$Be of 2.8 fm \cite{Tanihata85a,Zahar93}
and a rms halo radius of $^{14}$Be of about 5.5 fm
\cite{Tanihata85a} (both taken from Ref. \cite{Zahar93}). The
weighted mean rms matter radius of $^{14}$Be deduced in
Ref.~\cite{Ilieva2012} from several one-body density
parametrizations that are obtained by fitting the experimental
$p$+$^{14}$Be elastic scattering cross sections is 3.25(11) fm.

As can be seen, the theoretical results for the stripping and
diffractive processes have a similar shape. This is expected for
the energies considered in our work having in mind the results
obtained in Ref.~\cite{Hencken96} (see also
Refs.~\cite{Bona98,Aumann00,Yabana92,Yabana92,Anne93}) for
energies up to 100 MeV/nucleon.

The obtained values of the widths are 80.2 MeV/c and 77.8 MeV/c
with $R_{rms}=3.10$ fm and  $R_{rms}=3.50$ fm, correspondingly,
for the stripping reaction, and 115.7 MeV/c and 112.7 MeV/c for
the same values of radii in the case of diffraction process. These
values are in a reasonable agreement with the experimental width
($\Gamma= 92.2 \pm 2.7$ MeV/c) estimated in Ref. \cite{Zahar93}. A
quite weak dependence of the width at a given energy on the choice
of the rms radius was found. It turns out that the main condition
for the width to have a correct value is the parameters of the
potential well (e.g., of Woods-Saxon type) to provide the right
value of the binding energy of the pair of neutrons in the
$^{14}$Be nucleus. In our calculations the parameter values of the
Woods-Saxon potential are: $V_{0}=20.6$ MeV, $r_{0}=2.7$ fm,
$a_{0}=0.30$ fm for $R_{rms}=3.10$ fm and $V_{0}=16.8$ MeV,
$r_{0}=3.0$ fm, $a_{0}=0.40$ fm for $R_{rms}=3.50$ fm.

\section{Conclusions\label{s:conclusions}}

In the present work we followed two main aims. The first one was
to study elastic and quasielastic scattering of the neutron-rich
exotic nuclei $^{12}$Be and $^{14}$Be on $^{12}$C target at energy
56 MeV/nucleon, as well as their scattering on protons at 703.5
and 702.9 MeV/nucleon, correspondingly. The second aim was to
calculate the longitudinal momentum distribution of $^{12}$Be from
the fragmentation of $^{14}$Be on $^{12}$C at incident energy 56.8
MeV/nucleon.

In our hybrid model we calculate the real part of the optical
potential microscopically by the folding procedure in which
microscopic densities GCM and VMC for $^{12}$Be and GCM for
$^{14}$Be nucleus, as well as the symmetrized Fermi density (SF)
for both nuclei were used. Another ingredient of the folding
procedure is the effective $NN$ interaction related to the
$g$-matrix obtained on the basis of the Paris $NN$ potential. The
ReOP includes isoscalar and isovector direct and exchange
components. The ImOP is calculated microscopically as the folding
OP that reproduces the phase of the scattering in the high-energy
approximation. The free parameters of the model are the depths of
the real and imaginary parts of the OP. Their values are obtained
by fitting the experimental data on differential cross sections.
We calculated also the contributions of inelastic scattering to
the first 2$^+$ and 3$^-$ excited states in $^{12}$C in the
quasielastic $^{12,14}$Be + $^{12}$C processes. In addition, we
studied the role of the surface part $\rho^{(1)}_{SF}$ of the
density of the $^{12}$C target.

The main results from the work can be summarized as follows:

(i) In the case of the quasielastic $^{12}$Be+$^{12}$C scattering
all three densities of $^{12}$Be (SF, GCM, and VMC) give a
reasonable agreement with the data with the exception of the depth
and the position of the first minimum when only the elastic
channel is included. In the case of $^{14}$Be+ $^{12}$C an
agreement can be seen only at $\theta_{c.m.} > 5^\circ$.

(ii) The account for the contribution of inelastic scattering to
the first 2$^+$ state improves the depth of the first minimum and
leads to a left-shift correction of its place for both processes.
We note that a better agreement with the data for the
$^{12}$Be+$^{12}$C case for $\theta_{c.m.} < 8^\circ$ is obtained
by accounting only for the elastic scattering, while for larger
angles up to $17^\circ$ the elastic and inelastic scattering (to
the 2$^+$ state) give similar contributions and their sum allows a
reasonable agreement with the experimental data to be obtained.

(iii) The inclusion of the surface part $\rho^{(1)}_{SF}$ of the
$^{12}$C density leads to a correct reduction of the depth of the
first minimum and a good overall agreement with the data for the
$^{12}$Be+$^{12}$C case is achieved. However, for
$^{14}$Be+$^{12}$C this does not improve the agreement at angles
$\theta_{c.m.} < 5^\circ$.

(iv) A good agreement with the experimental $^{12}$Be+$p$ data for
the differential cross sections at 703.5 MeV/nucleon in the whole
range of angles is obtained with the use of SF and VMC densities
of $^{12}$Be. The use of SF density of $^{14}$Be leads also to a
very good agreement with the experimental $^{14}$Be+$p$ cross
sections data at 702.9 MeV/nucleon. The successful description of
both elastic scattering processes proves the important role of the
effects of relativization included in the calculations.

(v) In the second part of the work the longitudinal momentum
distribution of fragments in stripping and diffractive breakup
processes of $^{14}$Be nucleus on $^{12}$C is calculated in a
cluster model in which $^{14}$Be consists of $^{12}$Be core and a
$2n$ halo. OP’s of the interactions of $^{12}$Be and $2n$ with
the target are calculated within our hybrid model and their sum is
used in the folding procedure with the density corresponding to
the wave function of the relative motion of the clusters in
$^{14}$Be. Using the cluster OP’s the corresponding core ($c$)
and valence halo ($v=2n$) functions $S_c$ and $S_{v}$ (matrices)
are obtained within the eikonal formalism. They are used to
calculate the longitudinal momentum distributions of $^{12}$Be
fragments produced in the breakup of the halo-nucleus
$^{14}$Be+$^{12}$C at energy 56.8 MeV/nucleon. The obtained widths
are in a reasonable agreement with the experimental data and give
an important information for the halo structure of these nuclei. A
quite weak sensitivity of the computed widths to the choice of the
rms radius of  $^{14}$Be was found.

In general, we can conclude that our microscopic approach applied
to reaction studies with neutron-rich $^{12,14}$Be nuclei is
capable to reproduce the existing experimental data and allows to
support the two-neutron halo interpretation of these nuclei. More
definite conclusions about the relative role of the theoretical
ingredients of the microscopic model could be drawn when complete
and precise data from new reactions measurements, e.g., with the
novel generation of radioactive nuclear beam facilities, will
become available.

\section{Acknowledgments}
The authors are grateful to S.~C. Pieper for providing with the
density distributions of $^{12}$Be nucleus calculated within the
VMC method and to P.~Descouvemont for the density distributions of
$^{12,14}$Be nuclei obtained within the GCM. The authors thank
J.~J. Kolata for the discussion. Three of the authors (V.K.L.,
E.V.Z. and K.V.L.) thank the Russian Foundation for Basic Research
(Grant No.~17-52-18057 bolg-a) for the partial support. D.N.K.,
A.N.A., and M.K.G. are grateful for the support of the Bulgarian
Science Fund under Contracts No.~DFNI--T02/19 and No.~DNTS/Russia
01/3.

\end{document}